\documentclass[twocolumn]{article}
\usepackage{scicite}
\usepackage{times}

\usepackage{graphicx}

\def\kb{k_\mathrm{B}}
\def\YOg{\ensuremath{X^2\Sigma^+}}
\def\YOe{\ensuremath{A^2\Pi_{1/2}}}
\def\Gv{\mathbf{G}}
\def\Iv{\mathbf{I}}
\def\Sv{\mathbf{S}}
\def\Fv{\mathbf{F}}
\def\Nv{\mathbf{N}}

\newcommand{\ket}[1]{\left|#1\right\rangle}

\newenvironment{sciabstract}{%
\begin{quote} \bf}
{\end{quote}}

\newcounter{lastnote}
\newenvironment{scilastnote}{%
\setcounter{lastnote}{\value{enumiv}}%
\addtocounter{lastnote}{+1}%
\begin{list}%
{\arabic{lastnote}.}
{\setlength{\leftmargin}{.22in}}
{\setlength{\labelsep}{.5em}}}
{\end{list}}

\title{Magneto-optical trapping\\ of diatomic molecules} 

\author
{Matthew T. Hummon,$^{1}$ Mark Yeo,$^{1}$ Benjamin K. Stuhl,$^{1}$ \\Alejandra L. Collopy,$^{1}$ Yong Xia, $^{2}$ Jun Ye$^{1\ast}$\\
\\
\normalsize{$^{1}$JILA, National Institute of Standards and Technology and University of Colorado,}\\
\normalsize{ and Department of Physics, University of Colorado,}\\	
\normalsize{Boulder, CO 80309, USA.}\\
\normalsize{$^{2}$State Key Laboratory of Precision Spectroscopy, Department of Physics,}\\
\normalsize{East China Normal University, Shanghai,  China.}\\
\\
\normalsize{$^\ast$To whom correspondence should be addressed; E-mail: ye@jila.colorado.edu}
}

\date{}

\begin{document} 




\maketitle


\begin{sciabstract}

The development of the magneto-optical trap revolutionized the fields of atomic and quantum physics by providing a simple method for the rapid production of ultracold, trapped atoms.  A similar technique for producing a diverse set of dense, ultracold diatomic molecular species will likewise transform the study of strongly interacting quantum systems, precision measurement, and physical chemistry.   We demonstrate one- and two-dimensional transverse laser cooling and magneto-optical trapping of the polar molecule yttrium (II) oxide (YO).   Using a quasicycling optical transition we observe transverse Doppler cooling of a YO molecular beam to a temperature of 5~mK, limited by interaction time.   With the addition of an oscillating magnetic quadrupole field we demonstrate a transverse magneto-optical trap and achieve temperatures of 2~mK.  
\end{sciabstract}


Over the past quarter century, the magneto-optical trap (MOT) has been extended to two dozen atomic species \cite{Raab1987}.   This abundance of species makes ultracold atomic systems a powerful tool for studying a wide range of phenomena, from quantum-degenerate gases, physics beyond the Standard Model \cite{Guest_Radium}, and strongly correlated systems \cite{RevModPhys.80.885}, to applications in quantum information \cite{RevModPhys.82.2313} and simulation, quantum sensing, and ultraprecise optical clocks \cite{Swallows25022011}.   Ultracold polar molecules, with their additional internal degrees of freedom and complex interactions, yield even richer phenomena and open the door to new fields such as ultracold chemistry \cite{Carr:2009oz}.

Recently, many techniques have been developed for producing cold and ultracold samples of polar molecules.  Magneto-association and adiabatic transfer \cite{Ni:2008eq} of ultracold atoms can produce ultracold samples of polar molecules; though this technique is currently limited to bialkali species.  Buffer gas cooling \cite{Weinstein_CaH} and molecular beam slowing techniques, such as Stark \cite{Meerakker_Stark} and Zeeman \cite{Narevicius:08cg} deceleration, can produce molecules cold enough to load into conservative traps.  Further cooling of these trapped samples to temperatures below 10~mK via evaporative or sympathetic cooling remains technically challenging.  Optical cooling has been proposed \cite{Di-Rosa:2004kl,Stuhl:2008gb} and recently realized \cite{Shuman:2010hl}.  A MOT \cite{Stuhl:2008gb} would be the ideal tool for producing ultracold trapped samples of diatomic molecules, much as it is for atoms.

A MOT gains its utility by combining a spatially dependent trapping force with a fast dissipative cooling force.  With $1/e$ cooling rates on the order of $10^5$~s$^{-1}$, warm atoms can be cooled, trapped, and compressed in a few milliseconds over length scales of less than 1~cm.  To achieve these fast cooling rates, more than $10^4$ optical photons must be scattered at rates of more than $10^6$~s$^{-1}$, requiring a highly closed electronic transition.  The additional vibrational and rotational degrees of freedom present in molecules makes creating a closed cycling transition difficult in these systems.  Nevertheless, for certain molecules it is possible to construct quasicycling transitions with only a minimal increase in laser complexity \cite{Di-Rosa:2004kl,Stuhl:2008gb}. One-dimensional transverse laser cooling \cite{Shuman:2010hl} and longitudinal slowing \cite{Barry2012} of a SrF beam have been demonstrated. Opto-electric cooling of CH$_3$F molecules has resulted in temperatures as low as 29~mK \cite{Zeppenfeld:2012fk}. Here we demonstrate a new type of MOT, suitable for quasicycling transitions available in molecules.  We implement this system for the molecule yttrium (II) oxide (YO) and observe both viscous cooling and a magneto-optical spring force in a 2-D MOT.  

We begin by presenting the level structure of YO and describe the procedure for creating a quasicycling transition.  YO has a single naturally abundant isotopomer, $^{89}$Y$^{16}$O, and relatively simple hyperfine structure with nuclear spins $\Iv_\mathrm{Y}= 1/2$ and $\Iv_\mathrm{O}=0$.      The main cooling transition proceeds on $X^2\Sigma \rightarrow  A^2\Pi_{1/2}$ at 614~nm, as shown in Figure 1A.    The $A^2\Pi_{1/2}$ has a radiative lifetime of $\gamma^{-1} = 33$~ns \cite{Liu1977}, allowing for fast optical cycling. Diagonal Franck-Condon factors limit the vibrational branching of $A^2\Pi_{1/2}$  \cite{Bernard1983}.  Only two additional lasers at 648 and 649~nm  to repump the $v''=1$ and $v''=2$ vibrational levels are needed to limit vibrational branching loss to $< $10$^{-6}$.  The $\ket{A,v'; J' =1/2,F',+}$ manifold forms a highly closed transition with the $\ket{X,v''; N'' = 1,G'',F'',-}$~ manifold as shown in Fig.~1B, due to parity and angular momentum selection rules \cite{sci_methods}.   The magnetic field dependence of the ground manifold is shown in Fig.~1C.  Here, the ground electronic state is labeled by Hund's case $b_{\beta S}$ \cite{brownandcarrington} quantum numbers $\ket{X, v''; N'', G'',F'',p}$, with vibrational level $v''$, rotational level $N''$, intermediate quantum number $G''$ formed by coupling of electron and nuclear spin $\Gv = \Sv + \Iv$, and total angular moment $F''$  \cite{Childs1988}.  The excited \YOe~state is labeled by Hund's case $a$ quantum numbers, $\ket{A,v'; J',F',p}$, where $J'$ is the total electronic angular momentum, and $\mathbf{F  = J + I}$. The lowest rotational state $J' = 1/2$ is split into two states of opposite parity ($p = \pm$) separated by a $\Lambda$-doubling of $4.5$~GHz \cite{Bernard1983}.  This cooling scheme provides a rotationally closed transition at the expense of repumping of the $G''=0,1$ hyperfine ground states.   The repumping of hyperfine levels within each vibrational level can be achieved with a single laser by creating frequency shifted sidebands using an acousto-optic modulator\cite{sci_methods}.  In order to maintain optimal photon scattering rates for laser cooling, we destabilize optical dark states that form in the ground state Zeeman manifolds \cite{Berkeland:2002ad} by modulating the polarization of the cooling light between $\sigma^+$ and $\sigma^-$ with a voltage-controlled waveplate (Pockels cell). The modulation rate should be similar to the optical pumping rate, which in our case is on the order of several $10^6$~s$^{-1}$. 

This laser configuration for creating a quasicycling transition also lends itself naturally to creating a MOT.  Figure 1D shows the simplified YO level structure involved in the MOT.  In contrast to a typical atomic MOT, the ground state has a larger Zeeman degeneracy than the excited electronic state. A quadrupole magnetic field gradient provides a spatially dependent energy shift of the $G''=1$ ground state.  A molecule at position `r' (Fig.~1D) in the upper Zeeman level preferentially scatters photons from the laser propagating to the left due to selection rules and laser detunings.  This results in a restoring force towards the center of the trap.  Since we modulate the cooling light polarization to destabilize dark states (in the $G'' = 0$ manifold) we must also modulate the direction of the magnetic field in phase with the light polarization to maintain a restoring force for the MOT.  This is shown schematically in Figure 1D. Atomic MOTs with magnetic fields oscillating at 5~kHz can produce stable three-dimensional trapping \cite{Harvey:2008fk}.  Our modulation frequency of 2~MHz is set by the optical pumping rate.

To fully describe the cooling and trapping forces for YO would involve 44 molecular levels, 15 optical frequencies with time-dependent polarization, and a time-dependent magnetic field.  Such a calculation is beyond the scope of this report.  Nevertheless, a simple multi-level rate equation model \cite{Kloeter:2008im} can be used to extend the results from the two-level models to provide physical insight into the observed dynamics of a YO MOT.  In the limit of small laser detunings and low laser power, the result from the two-level system for the Doppler and magneto-optical cooling force in one dimension can be expressed in the form \cite{Metcalf:1999la}:
\[ F  = -\beta v- \kappa r \]
\[ \beta  = \frac{-8 \hbar k^2 \delta s_0}{\gamma (1+s_0 + (2\delta/\gamma)^2)^2}\]
 \[ \kappa  = \mu' A\beta/ \hbar k \]
Here, $F$ is the total force experienced by the molecule,  $\beta$ characterizes a viscous drag (Doppler) force, proportional to the molecule's velocity, $v$, and $\kappa$ represents a magneto-optical spring force proportional to the molecule's displacement, $r$, from the magnetic field minimum.  The forces are parameterized by the wavenumber, $k$, of the cooling transition, the detuning, $\delta$, of the cooling laser from the molecular transition, the resonant saturation parameter $s_0$, the differential magnetic moment between ground and excited state $\mu'$, the magnetic field gradient $A$, and reduced Planck's constant $\hbar$.  

For multi-level systems with $N$ ground states and a single excited state the optimal (maximum) damping parameter $\beta_N$ scales roughly as $\beta_N\sim2\beta_1/(N+1)$, and is achieved when all grounds states are driven. This results from lower photon scattering rates due to decreased population in the excited state from the additional degeneracies of the $N$ ground states.  Despite the slower damping rates, the Doppler cooling limit for the multi-level system remains unchanged from the two-level result, $T_\mathrm{dop} = \hbar \gamma/2\kb = 116~\mu$K.   The slower damping rates are balanced by slower heating rates from photon recoils.  For our time-dependent magnetic field, it suffices to replace the two-level system $\kappa$ with a time averaged $\bar{\kappa}$.  Averaging the magneto-optical force over a single cycle yields $\bar{\kappa} = (2 \sqrt{2}/\pi) \cos(\phi)\kappa$, where $\kappa$ is calculated with the root-mean-square (RMS) field gradient, and $\phi$ is the phase between the modulations of the MOT field and the optical polarization.  By changing the relative phase $\phi$,  we can change the magnitude and sign of the MOT spring force independently of the Doppler force.  

To characterize the MOT we use a cryogenic buffer gas molecular beam apparatus \cite{Maxwell:2005ul}, depicted in Figure 2.  The YO molecules are produced via laser ablation of a sintered Y$_2$O$_3$ pellet located inside a copper cell filled with a 4.5~K helium buffer gas.  The YO molecules thermalize translationally and rotationally via collisions with the 4.5~K helium buffer gas.  In-cell laser absorption measurements indicate initial $\ket{X, 0;N''=1}$ densities on the order of $10^{10}$~cm$^{-3}$, corresponding to more than $10^{10}$ molecules produced per ablation pulse.  A YO molecular beam is formed by extraction of the molecules through a 3~mm diameter aperture in the side of the buffer gas cell.  The molecular beam is collimated by a second aperture, 2.5~mm in diameter, placed 130~mm from the buffer gas cell.  The direction of propagation of the beam is defined as the $z-$axis.    This results in a molecular beam with longitudinal velocity of $v_z \sim120$~m/s, with full-width-half-maximum (FWHM) spread of 40 m/s in the longitudinal velocity distribution ($T_z\sim3.3$~K), as determined by Doppler shift fluorescence spectroscopy.  The transverse temperature of the molecular beam after the collimating aperture is $T_\perp \sim25$~mK.

Following the collimating aperture the molecular beam travels to a 10~cm long interaction region, where the molecules interact with the cooling lasers.  The cooling lasers have a FWHM beam diameter of $\sim3$~mm and make 11 round-trip passes through the interaction region to provide a molecule-laser interaction time of $t_\mathrm{int} \sim 275~\mu$s.  The propagation direction of the cooling lasers defines the $x-$axis. Additionally, the magnetic field coil used for the MOT has a rectangular baseball coil geometry, with dimensions of 5 x 5 x 15 cm.  The coil consists of 25 turns of Litz wire in series with a tuning capacitor, forming an LC-resonator with resonant frequency $\omega_{0} = 2\pi \times2$~MHz and quality factor $Q = 77$.  Power is coupled into the MOT coils via a transformer designed to impedance match the MOT coil to a $50~\Omega$ transmission line.  Assuming perfect coupling, a driving power of 20~W corresponds to RMS field gradients of $A = 6$~gauss/cm.  The MOT field coil is located inside the vacuum chamber and is set in a low-outgassing, thermally conductive, electrically insulating epoxy and water cooled to avoid overheating of the coils during operation.  We monitor the phase of the MOT field with a pick-up coil and phase lock the MOT field to the polarization modulation signal \cite{sci_methods}.

Following the interaction region the molecules traverse a 30~cm long region for ballistic expansion.  Finally, the molecules enter a probe region where they are first optically pumped back into the ground vibrational state with a multi-pass ``clean-up'' beam consisting of $v=1,2$ repump lasers.  A retro-reflected probe beam, derived from the same laser beam used for the cooling transition, propagates along the $x-$axis and interrogates the molecules. The cycling fluorescence from the probe beam is collected along the $y-$axis and imaged onto a CCD camera.  To extract transverse temperatures from the molecular beam image we fit the image  profile along the $x-$axis to an expected functional form calculated from a Monte Carlo simulation of the molecular beam. 

Figure 3A shows transverse molecular beam profiles under various conditions.   Curve (i) in Fig.~3A shows the unperturbed molecular beam (cooling lasers off), with a transverse temperature of 25~mK.   If only the $v''=0$ cooling laser is turned on, the YO molecules are efficiently pumped into $v''=1,2$ levels, leaving a depleted beam signal shown as curve (ii).   The curves (iii) and (iv) in Fig.~3A correspond to the YO molecules under the presence of cooling lasers detuned by $\delta/2\pi = -5$ and $+5$~MHz, respectively. We define the detuning, $\delta = \omega_\mathrm{laser} - \omega_\mathrm{YO}$, as a uniform detuning of all the $v''=0$ cooling lasers from their respective transitions, with $\delta =0$ corresponding to no observable change in temperature. The $v''=1,2$ repump lasers remain on resonance.  Cooling of the molecular beam is observed as a narrowing of the molecular beam profile and increase in the number of molecules at the center of the beam.  As expected, for small negative (positive) detunings we observe cooling (heating) of the molecular beam.  Figure 3B shows the Doppler force dependance over a range of laser detunings.  The observed temperatures agree well with a simulation of the force.  Due to the closely spaced hyperfine levels in the $G''=1$ manifold of the ground state, a laser with $\delta/2\pi = +10$~MHz for the $G''=1, F''=0$ level will act as a negatively-detuned beam for the $G''=1, F''=1,2$ levels.  The simulation of the Doppler force indicates that this effect is responsible for the cooling observed at large positive detunings.  Additionally, we have negatively detuned the laser for each ground hyperfine manifold independently, while leaving the remaining $v''=0$ hyperfine lasers on resonance, and observed cooling in each case.  This indicates that all hyperfine manifolds in the ground state participate in cooling.  Figure 3C shows the dependance of the Doppler cooled temperature on polarization modulation rate for a YO molecular beam of initial transverse temperature of $15$~mK and laser detuning $\delta = -5$~MHz.  The cooling and photon scatter rate is clearly limited by Zeeman dark states for polarization modulation rates less than 5~MHz.  

The final Doppler cooled beam temperatures are in the range of 5 to 10~mK, well above the Doppler limit of $116~\mu$K.  The simulations of the Doppler cooling force indicate this is due to a finite interaction time and cooling rate.  The final temperature of the beam can be expressed as $T_\mathrm{f} = T_\mathrm{i} \times \exp[- t_\mathrm{int} \Gamma_\mathrm{D}]$, where $T_\mathrm{i} = 25$~mK is the initial beam temperature, and the Doppler cooling rate $\Gamma_\mathrm{D} = (2\beta/m)$ where $m = 105$ AMU is the mass of YO.  This implies an experimental value of $\Gamma_\mathrm{D} \sim 5 \times 10^3$~s$^{-1}$, which is in good agreement with the cooling rate $\Gamma_\mathrm{D} = 8 \times 10^3$~s$^{-1}$ predicted by the multi-level rate equation model.

Under optimal cooling conditions we observe that $85\%$ of the molecules remain after cooling.  This implies that the branching ratio into dark states is on the order of $\eta \sim 10^{-4}$.  One possible loss channel is via decay to the $A'^2\Delta_{3/2}$ electronic state, shown in Fig.~1A. A calculation of the dipole transition strength for the  $A^2\Pi_{1/2} \rightarrow A'^2\Delta_{3/2}$ transition \cite{Langhoff1988} indicates that the branching fraction could be as large as $\eta = 4\times 10^{-4}$.  Molecules that decay into $A'^2\Delta_{3/2}$ will subsequently decay into the ground $X^2\Sigma$, $N'' =0,2$ levels \cite{Simard1992}, acquiring a parity flip from emitting an additional photon.  This may be a useful method for accumulating molecules in the ground rotational state.  Alternatively the $X^2\Sigma \rightarrow A'^2\Delta_{3/2}$ transition at 690~nm could be used for narrow-line cooling \cite{Rehbein:2007zt} or to pump the molecules back into the $N''=1$ level by applying a weak electric field to mix parity states in the $A'^2\Delta_{3/2}$ level.  

With the Doppler cooling process well characterized, we can now turn to the MOT's properties.  Figure 4A shows a typical molecular beam image after passing through the 1-D MOT.  Curve (i) shows the unperturbed molecular beam.  Curves (ii) and (iii)  show the molecular beam after passing through the 1-D MOT for relative MOT phase $\phi = 0^\circ$ (trapping) and $\phi = 180^\circ$ (anti-trapping), respectively.  While both curves (ii) and (iii) exhibit cooling, curve (ii) clearly shows the enhancement of molecules at the center of the beam due to the MOT spring force.  Figure 4B  shows the dependance of the final temperature of the molecules on the relative MOT phase, $\phi$.  The observed dependance agrees well with a Monte Carlo simulation of the MOT, yielding a value of the MOT oscillation frequency of $\omega_\mathrm{MOT} = \sqrt{\bar{\kappa}/ m} \sim 2\pi \times 155$~Hz.  This agrees well with the predicted value of $\omega_\mathrm{MOT} \sim 2\pi \times 160$~Hz, derived from the experimentally measured value of $\beta$ and calculated magnetic field gradient $A_\mathrm{RMS} = 6$~gauss/cm.  Even though the molecules traverse the magneto-optical trap in a fraction of a trap oscillation, the molecular beam intensity enhancement due to the MOT is clear and can be well explained by the optical pumping rate model.  

Figures 4C and 4D compare the observed temperatures for Doppler cooling and the MOT for varying  cooling and repump powers, respectively.  Simulations of the cooling force are shown as a solid line. In Fig.~4C we note that at large cooling powers, the Doppler cooling rate decreases (larger final temperature) due to power broadening of the cooling transition.  Although our simple model also predicts this effect for the MOT, we observe the coldest MOT temperatures for the largest cooling power.   This disagreement is likely due to the more complicated level structure in the MOT that is not accounted for in the simulation.  In Fig.~4D, at low repump power, the cooling rate for both Doppler cooling and the MOT is limited by the repump time out of the $v''=1,2$ levels, while at large repump power, the cooling rate saturates. 

While the 1-D MOT provides simple pedagogical understanding, to enhance the brightness of the molecular beam we have implemented Doppler cooling and the MOT in 2-D.  The cooling lasers then propagate along both the $x$ and $y$ directions.  The action of the MOT along the $y$-direction directly resulted in an increase of detected molecules.  The observed value of $\beta_x$ in 2-D is roughly half that observed in 1-D \cite{sci_methods}.  This is consistent since half the photons are now scattered along the $y$-axis and do not provide any cooling in the $x$-direction.  We also note that for cooling in dimensions of 2 or more, modulation of the magnetic field and polarization may not be necessary to remix hyperfine states, simplifying the magnetic coil design.   Type II atomic MOTs operate with similar level structures, and remixing occurs from lasers propagating in orthogonal directions \cite{Oien:1997ab}.  

From our observations of the 1-D and 2-D MOT, we can estimate the properties of a 3-D MOT based on this system.    We first define a generic 1-D damping rate, $\Gamma = \beta/m$.  It is straightforward to show that the phase space density of molecules in a 3-D MOT will evolve as $\rho \propto n T^{-3/2} \propto  \rho_0 \exp[ (3\Gamma/ 2 - \eta\gamma_p )t]$, where $\eta$ is the branching ratio into dark states and $\gamma_p$ is the photon scattering rate. Phase space increase requires $\eta < 3 \Gamma/ (2\gamma_p) \sim 10^{-3}$, where we have substituted typical parameters for YO, $\Gamma = 2\times 10^3$~s$^{-1}$, and $\gamma_p = 3\times10^6$~s$^{-1}$.  The lifetime of molecules in the MOT will be limited by optical pumping into dark states, with $\tau_\mathrm{MOT} \sim (\eta \gamma_\mathrm{p})^{-1} > 1$~ms for $\eta < 4\times 10^{-4}$. Typical capture velocities will be on the order of several m/s.  Loading of a 3-D MOT could be achieved either from a slow molecular beam \cite{Barry2012,Slowbeam} or possibly directly from molecules produced inside a buffer gas cell.  Thus, a clear path exists to a 3-D MOT that produces cold, dense samples of diatomic molecules. 

\bibliography{/Users/matt/Dropbox/articles/articles}

\bibliographystyle{Science}

\begin{scilastnote}
\item We acknowledge funding support for this work from AFOSR (MURI), DOE, NIST, and NSF PFC at JILA.  M.T.H. acknowledges support from the NRC.  We thank K. Cossel and F. Adler for technical help.
\end{scilastnote}

\clearpage

\onecolumn

 \begin{figure}[h]
 \includegraphics[width=4.75in]{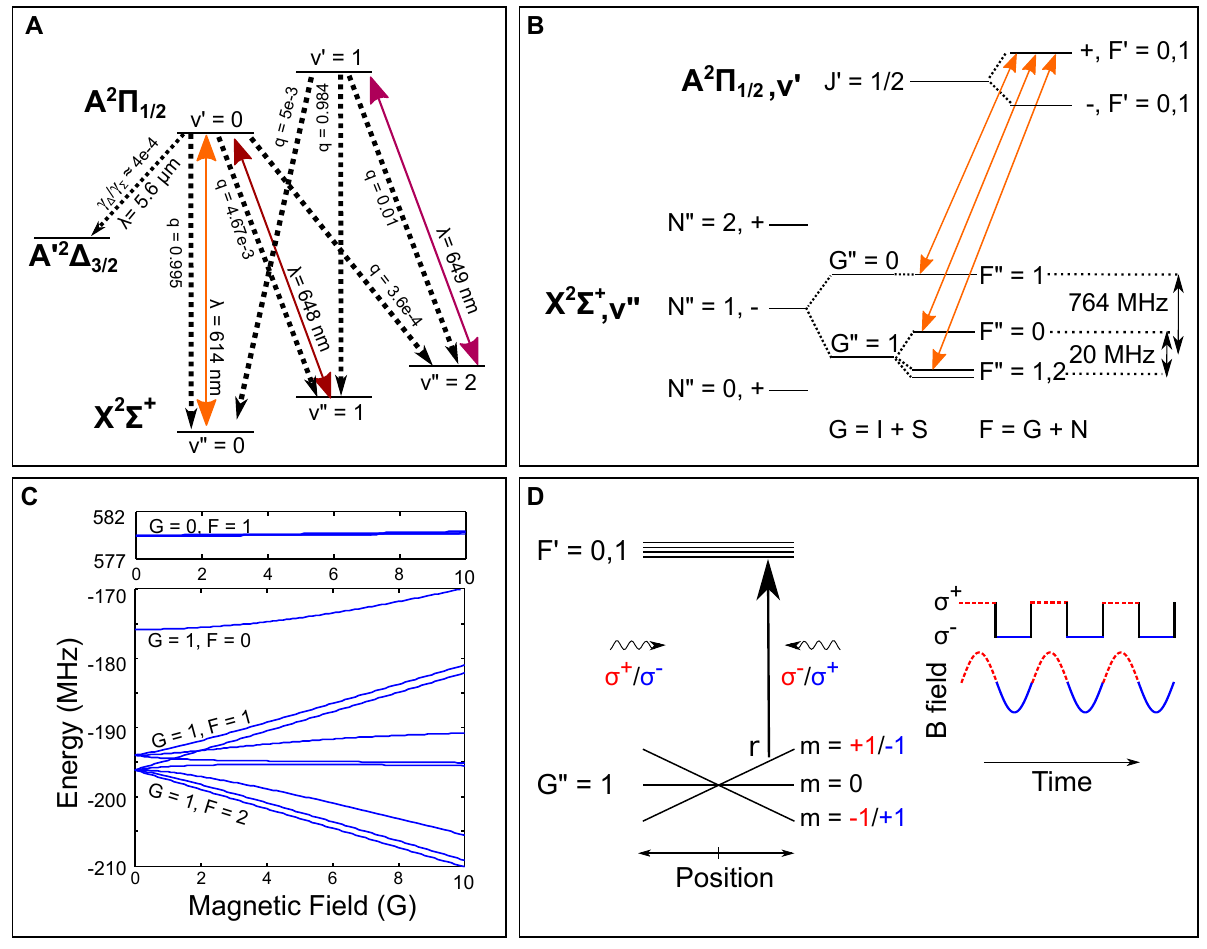}%
 \caption[YO Level Structure ]{(A) YO vibronic structure.  Dashed arrows indicate decay paths with corresponding Franck-Condon factors, $q$ \cite{Bernard1983}.  Solid arrows indicate cooling and repump laser transitions. (B) Rotational and hyperfine structure of the $X$ and $A$ states.  Solid arrows indicate the three hyperfine pumping components used in this work. (C) Zeeman structure for the \YOg, $N''=1$ state. (D)  Schematic of the MOT level structure and the modulation waveforms for optical polarization and magnetic field.}
\label{fig:YOlevels}
 \end{figure}

\begin{figure}[h]
 \includegraphics[width=4.75in]{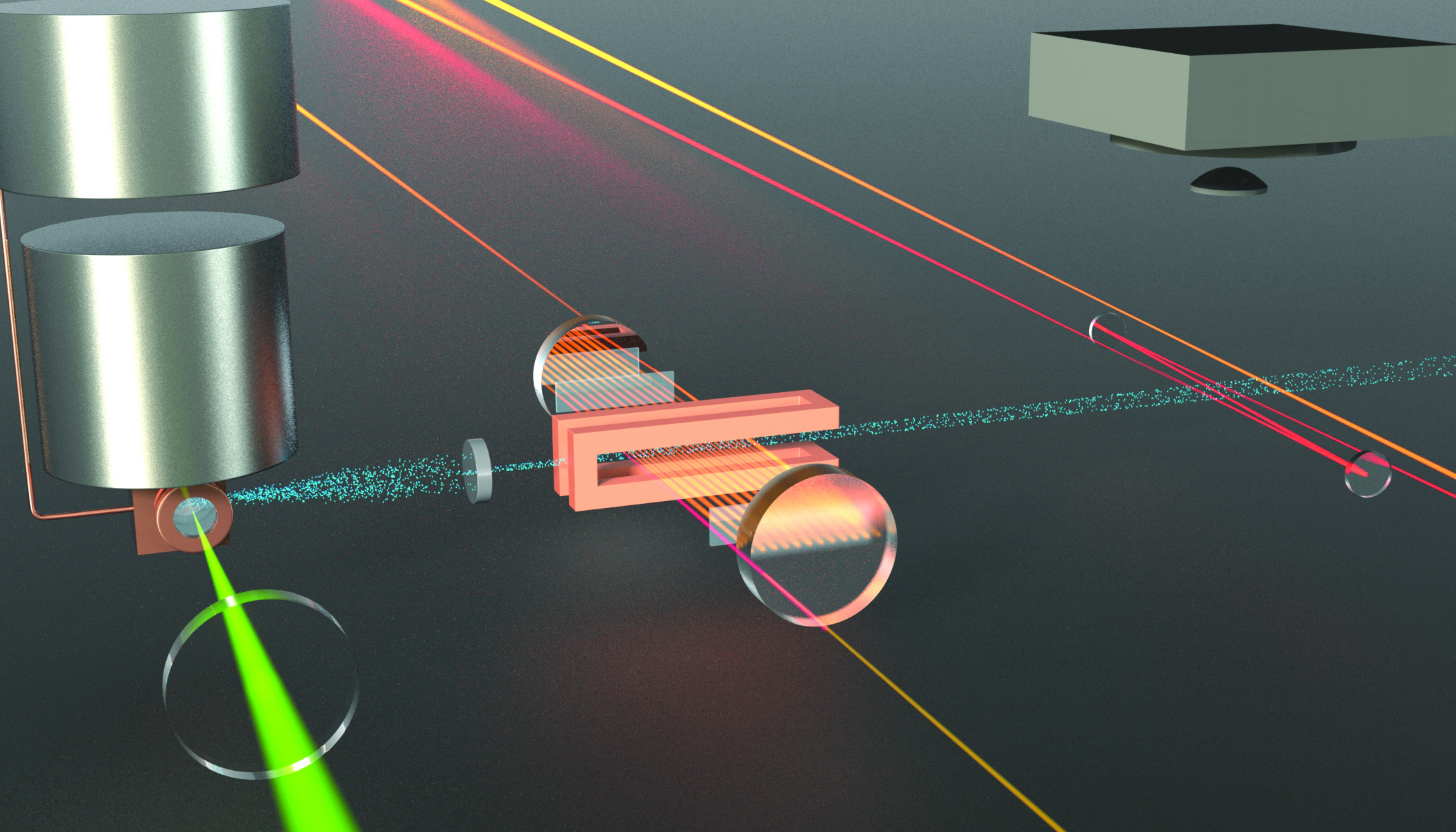}%

\caption[Beam Apparatus]{Depiction of the MOT apparatus, shown in its 1-D implementation for clarity.  The 2-D system has cooling laser beams propagating in the vertical directions as well.  YO molecules are produced via laser ablation (green) inside a cryogenic buffer gas cell, shown at left.  The YO molecular beam (shown blue) is collimated by an aperture and then passes through the cooling region, shown at center.  The cooling region consists of a rectangular magnetic field coil and a multipass laser beam interaction region.  The multipass consists of a pair of mirrors and $\lambda/4$ waveplates to provide the correct polarization of light for the MOT over the 10~cm interaction length.   After passing through the cooling region,  the YO molecules are optically pumped into the vibrational ground state (red beams) and the molecular beam is imaged using resonant fluorescence and a CCD camera, shown at right. Image credit: Brad Baxley.} 
\label{fig:Apparatus}
\end{figure}

\begin{figure}[h]
 \includegraphics[width=4.75in]{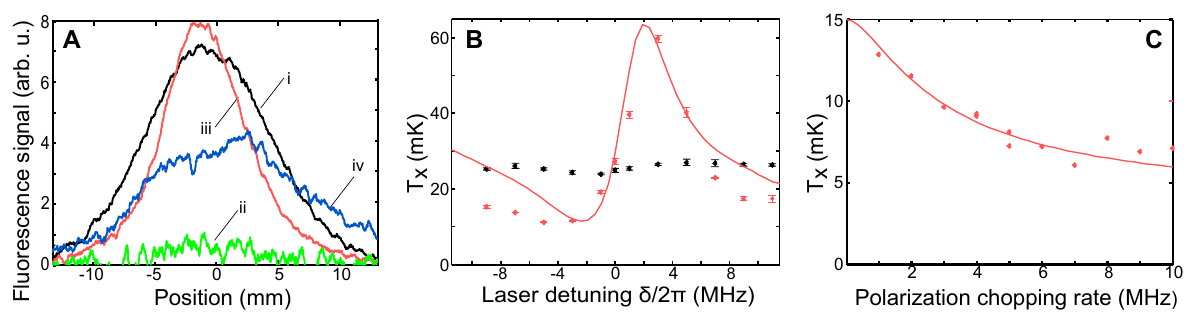}%

\caption[YO doppler cooling]{1-D Doppler cooling.  (A) Molecular beam profiles for (i) an unperturbed beam, (ii) depletion due to lack of vibrational repumping, (iii) Doppler cooling and (iv) Doppler heating. (B) Transverse temperature vs. cooling laser detuning, $\delta$, for unperturbed (black) and Doppler cooled (red) beams. (C) Transverse temperature vs. polarization modulation rate with an initial beam temperature of 15~mK.  The solid lines in (B) and (C) represent simulations using a multi-level rate equation model for the Doppler cooling force.}
\label{fig:Doppler}
\end{figure}

\begin{figure}[h]
\includegraphics[width=4.75in]{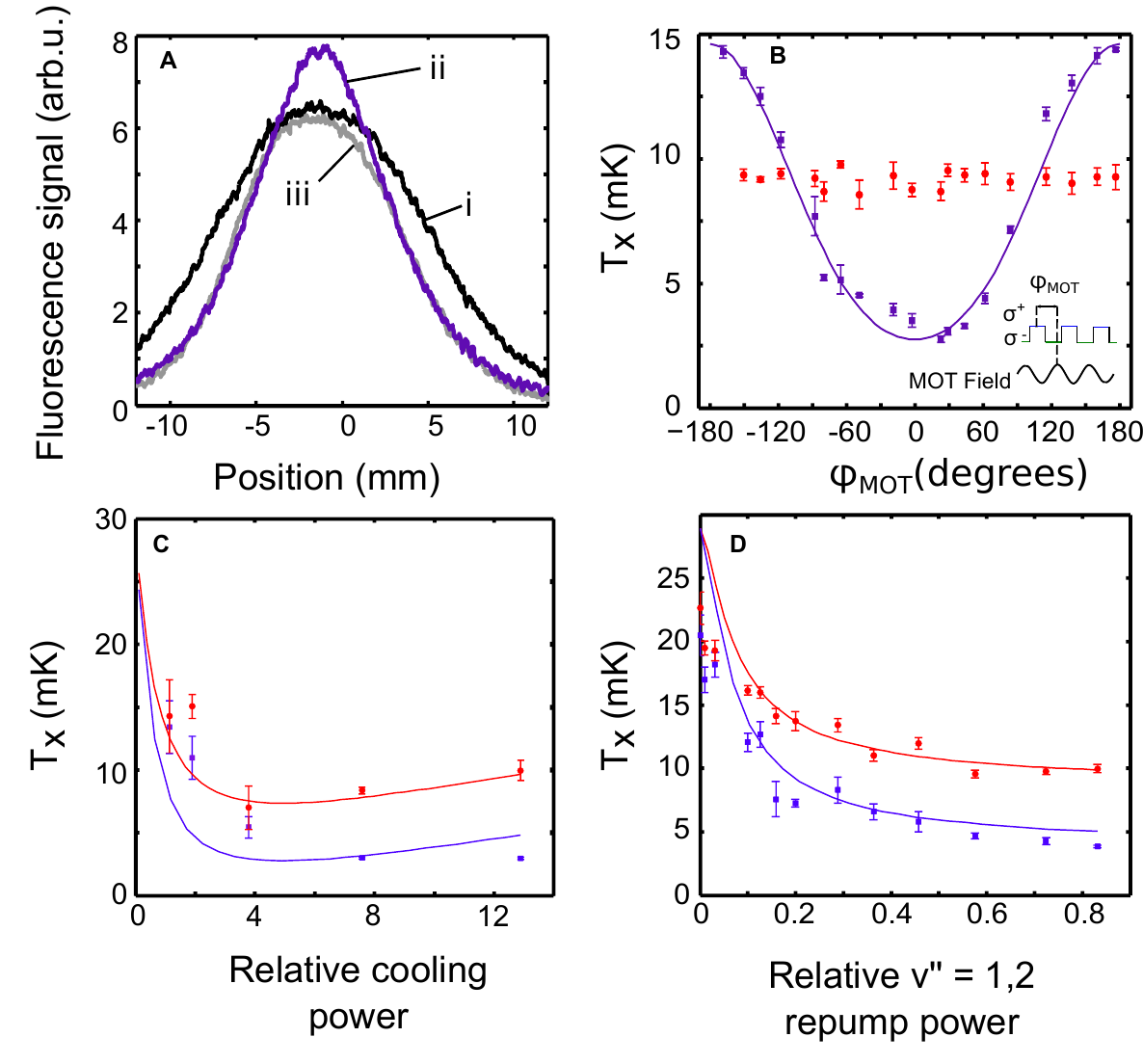}%

\caption[YO MOT]{(A) Molecular beam profiles for (i) an unperturbed beam, (ii) a 1-D trapping ($\phi=0$) MOT, and (iii) an antitrapping $\phi = 180^\circ$ MOT. Comparisons of transverse temperature for Doppler cooled (red) and MOT cooled (purple) beams, varying (B) relative MOT phase, (C) relative cooling power, and (D) relative repump power.  Simulations of the cooling forces are shown with solid lines.}
\label{fig:MOT}
\end{figure}

\clearpage

\section*{Supplementary Materials}

\paragraph{Optical cycling transition}

Figure 1B shows the rotational and hyperfine structure of the ground \YOg~state of YO, which is of Hund's case $b_{\beta S}$ \cite{brownandcarrington}.    The ground electronic state can be labeled by quantum numbers $\ket{X, v''; N'', G'',F'',p}$, where $v''$ indicates vibrational level and $N''$ indicates rotational level.  A strong Fermi contact interaction couples the electronic spin $\Sv = 1/2$ to the nuclear spin $\Iv =1/2$ of the $^{89}$Y to form an intermediate quantum number $\Gv = \Sv + \Iv$.  The electron spin-rotation interaction then couples molecular rotation $\Nv$ to form total angular momentum  $\Fv = \Gv+\Nv$.  The rotational states have parity, $p = (-1)^{N''}$, denoted by $+/-$ in Fig \ref{fig:YOlevels}(b).   The excited \YOe~state is labeled by Hund's case $a$ quantum numbers, and can be labeled by quantum numbers $\ket{A,v'; J',F',p}$, where $J'$ is the total electronic angular momentum, and $\mathbf{F  = J + I}$. The lowest rotational state $J' = 1/2$ is split into two states of opposite parity separated by the $\Lambda$-doubling of $4.5$~GHz \cite{Bernard1983}.  Each parity state consists of a pair of hyperfine manifolds $F' = 0,1$ that are not spectroscopically resolved.   

 Parity selection rules require the $\ket{A, v'; 1/2,F',+}$ state to decay to a negative parity ground state, implying $N''$ must be odd.  Each ground level $\ket{X,v'';N'',G'',F''}$ can be expressed as a superposition of states expressed in the $\ket{X, v''; J''= N''\pm1/2, F''}$ basis.  The angular momentum selection rule $|J''-J'| = 0,1$ then only allows decays to the  $\ket{X,v'',N'' = 1,G'',F'',-}$ state.  Mixing of the $\ket{X,v''; 1,1, 2}$ with the $\ket{X,v''; 3,1,2}$ level  or the $\ket{A,1/2,1}$ with the $\ket{A,3/2,1}$ level via the hyperfine interaction can lead to branching loss ratios on the order of $(10~\mathrm{MHz}/ 30~\mathrm{GHz})^2 < 10^{-6}$ and can be ignored.   
 
One additional complication arises when simultaneously driving the optical transitions from the energy degenerate sub-levels within the hyperfine manifolds.  In particular, for transitions of the the type $F''\rightarrow F'=F''$ or $F''-1$, where $F''$ is an integer, optical dark states will arise for any choice of static laser polarization \cite{Berkeland:2002ad}.  In these dark states, the ground molecular hyperfine state is no longer coupled to the excited state by the optical cooling laser, and the cooling  ceases.  The common solution is to destabilize the dark states by making them time dependent, either by applying an external magnetic field to make the energy levels non-degenerate, or by modulating the polarization of the driving optical field.  While application of a magnetic field is straightforward experimentally, for this level structure in YO it does not work.  The $\ket{X,v'';1,0,1}$ state is a spin singlet state, and the energies of its Zeeman sublevels are insensitive to application of an external magnetic field, as shown in the upper panel of Fig.~1C.  Therefore, we instead modulate the polarization of the cooling light between $\sigma^+$ and $\sigma^-$.  In order to maintain optimal photon scattering rates for laser cooling, the modulation rate should be similar to the optical pumping rate, which in our case is on the order of several $10^6$~s$^{-1}$.  This is achieved experimentally with a voltage-controlled waveplate, also know as a Pockels cell.

\paragraph{Laser cooling setup}

To produce the light necessary for creating a cycling transition in YO we use a setup consisting of three lasers, as shown in Figure S1.   The laser light for the main cooling transition at $\lambda =  614$~nm is generated from a ring cavity dye laser.  The dye laser frequency is stabilized to the $\ket{X, 0; 1,0,1}\rightarrow\ket{A,0; 1/2, F', +}$ transition frequency.   Cooling light for the $\ket{X, 0; 1,1, F''}$ manifold is generated by frequency shifting a portion of the main cooling beam using an acousto-optic-modulator (AOM) operated at two frequencies, 760 and 774~MHz.  The $v''=1$ and $v''=2$ repump light at 648~nm and 649~nm is generated from a pair of master external cavity diode laser (ECDL) and injection locked slave laser setups.  The 3~mW output from a master ECDL is stabilized to the corresponding $\ket{X,(1,2); 1,0,1}\rightarrow \ket{A, (0,1); 1/2, F', +}$ transition.  This master ECDL laser is used to injection lock a high power laser diode ($\sim80$~mW).   Most of the slave diode power is used for repumping of the $\ket{X, (1,2); 1,0,1}$ level, while a portion ($<10$~mW) is frequency shifted using a double pass AOM and used to injection lock a second high power diode for repumping the $\ket{X, (1,2); 1,1,F''}$ manifold.  To address all of the $F''=0,1,2$ manifolds in the $G''=1$ state, the AOM frequency is modulated at a rate of 1~MHz to provide a frequency shift between 760~MHz  and 774~MHz.  For each vibrational level $v''=0,1,2$, the $G''=0$ and $G''=1$ beams are combined on separate polarizing beam splitter cubes.  The $v''=1,2$ repump light is then combined on a non-polarizing $50-50$ beam splitter cube, with half of the power for the cooling beam and half of the power for the ``cleanup''  repump beam.  The $v''=1,2$ repump beam is combined with the $v''=0$ cooling beam using a dichroic mirror.  The cooling beam at this point consists of frequencies for $v''=0,1, 2$ levels with the beam for each level consisting of a pair of beams with orthogonal linear polarizations, for $G''=0$ and $G''=1$ manifolds.   This beam then passes through the Pockels cell allowing the polarization of the beams to be modulated between $\sigma^+$ and $\sigma^-$ at rates of up to $10$~MHz.  

We frequency stabilize the cooling and repump lasers via optical heterodyne measurements with light from a self-referenced octave-spanning erbium-doped fiber frequency comb. 

\paragraph*{Magneto-optical trap phase stabilization}
During typical experimental conditions, the MOT coils are turned on and off, leading to cycling of the coil temperature and drifts of the resonant frequency, $\omega_\mathrm{0}$.   Since the MOT is driven at a fixed excitation frequency,  drifts of the resonant frequency will lead to drifts in the phase shift of the MOT field.  To stabilize this phase shift, we monitor the phase of the MOT field using a pickup coil and phase lock it to the polarization modulation reference signal using the circuit diagram shown in Figure S2a.  Figure S2b shows the final MOT temperature versus the drive power of the MOT coils.  The exact dependence of the final MOT temperature on magnetic field gradient is complicated and not fully understood.

\paragraph*{Comparison of 1-D and 2-D MOT}

Figure S3 shows a comparison of the MOT operation in one and two transverse dimensions.  The total available cooling laser power is split equally between laser beams that propagate along the $x-$ and $y-$axes.  Figure S3A shows the molecular beam profile along the $x-$axis for $1-$D cooling lasers propagating along the $x-$axis.  Figure S3B shows the molecular beam profile along the $x-$axis for $2-$D cooling lasers propagating along both the $x-$ and $y-$axes.  The 1-D MOT exhibits a faster cooling rate along the $x-$axis than the 2-D MOT, as shown by the molecular beam's narrower width.  This arises from the finite photon scattering rate.  In two dimensions, the photons scattered by the molecules from the laser propagating along the $y-$axis do not lead to cooling along the $x-$axis.  They do, however, lead to cooling and compression along the $y-$axis, which results in more molecules in the field of view of the probe beam (propagating along the $x-$axis).  Thus, the 2-D MOT produces a brighter molecular beam, as shown by the larger molecule number in Figure S3B.   

\clearpage

\setcounter{figure}{0}
\renewcommand{\thefigure}{S\arabic{figure}}

\begin{figure}[h]
\includegraphics[width=4.75 in]{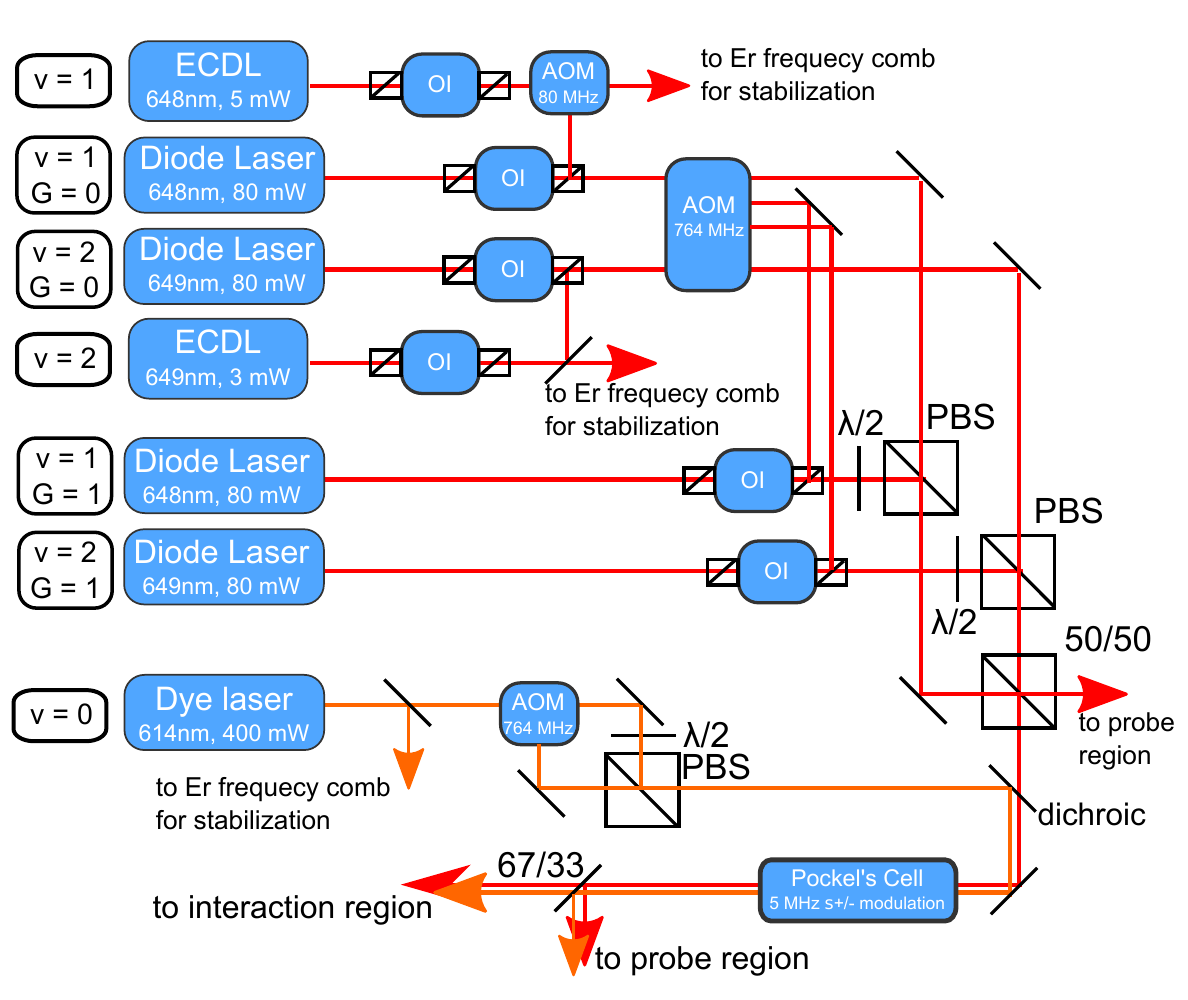}
\caption{Laser cooling setup for YO.  Abbreviations: ECDL, external cavity diode laser;  OI, optical isolator; AOM, acousto-optic modulator; $\lambda/2$, half-wave retardation plate; PBS, polarizing beam splitter.  Due to the limited output power of the ECDL, we use injection seeding to boost the usable optical power for the MOT experiment.}
\end{figure}

\begin{figure}[h]
\includegraphics[width=4.75 in]{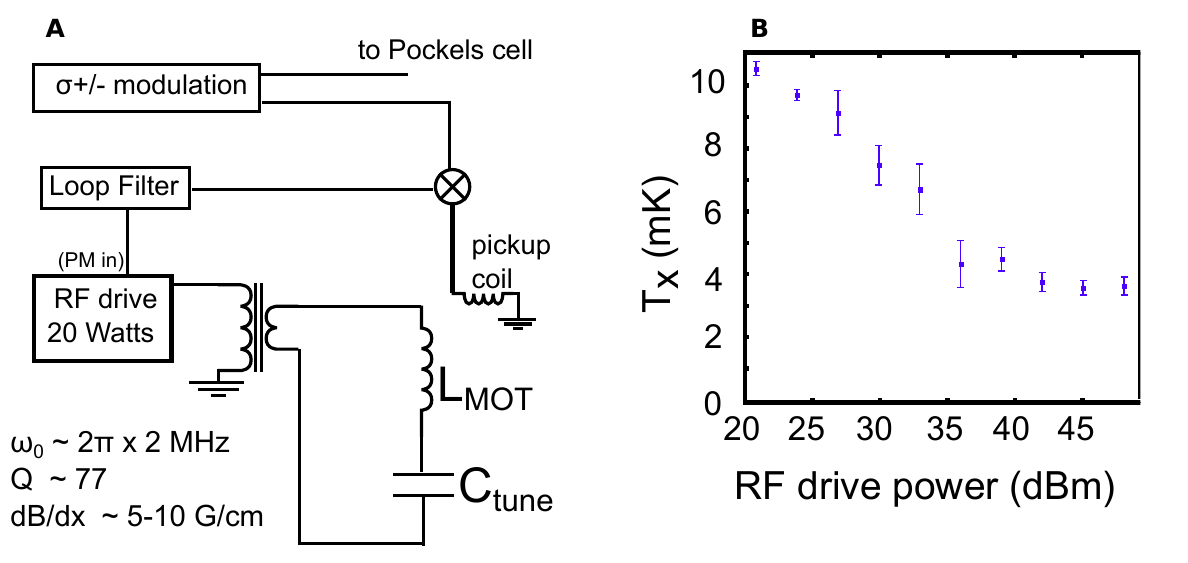}
\caption{(A) Circuit diagram for the MOT coil.  The phase of the MOT is monitored via a pickup coil.  The signal from the pickup coil is mixed with the reference signal from the polarization modulation and used to phase lock the MOT coil to the polarization modulation. (B) The measured molecular MOT temperature as a function of RF drive power, which corresponds to the MOT magnetic field gradient under modulation.}
\end{figure}

\begin{figure}[h]
\includegraphics[width=4.75in]{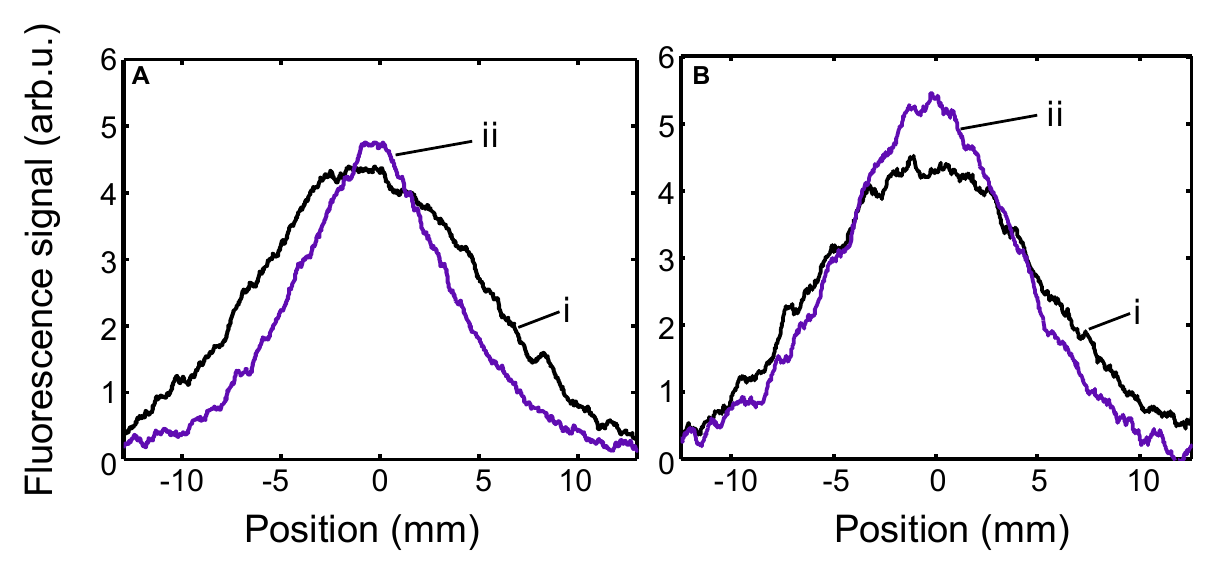}
\caption{Comparison of 1-D and 2-D MOTs.  (A) Molecular beam profiles for an  (i) unperturbed beam and (ii) 1-D MOT.  (B) Molecular beam profiles for an  (i) unperturbed beam and (ii) 2-D MOT.  The 1-D MOT exhibits a faster cooling rate along the observed axis, while the 2-D MOT produces a brighter molecular beam. }
\end{figure}


\end{document}